\documentstyle[aps,pra,epsf,multicol]{revtex}
\begin{document}
\draft
\title{Proposal for teleportation of an atomic state via cavity decay}
\author{S. Bose $^2$, P. L. Knight $^1$, M. B. Plenio $^1$ and  V. Vedral $^2$}
\address{$^1$Optics Section, The Blackett Laboratory, Imperial College, London SW7 2BZ, England}
\address{$^2$ Centre for Quantum Computing, Clarendon Laboratory,
	University of Oxford,
	Parks Road,
	Oxford OX1 3PU, England}

\maketitle
\begin{abstract}
We show how the state of an atom trapped in a cavity can be teleported to
a second atom trapped in a distant cavity simply by
detecting
photon decays from the cavities. This is a rare
example of a decay mechanism playing a constructive role in
quantum information processing. The scheme is comparatively easy to
implement, requiring only the ability to
trap a single three level atom in a cavity.
 
\end{abstract}


\begin{multicols}{2}

 Spontaneous decay is popularly regarded as a coherence loss mechanism 
in a 
quantum system. As such, one may not expect such a process to 
be helpful in quantum information processing \cite{ple1}. Two recent papers
\cite{cab,ple} tend to dispel this myth by showing how the detection (or the non detection) of decays can be used to entangle the states of distinct atoms. 
Here, we 
show that the above approach is not limited to the establishment of
entanglement, but can actually be used for {\em genuine} quantum information
processing such as {\em teleportation} \cite{ben}.
In our proposal, the
states to be teleported (the "stationary qubits") are atomic states,
ideal for the storage of 
quantum information. Quantum information is
physically transferred from place to place 
via photonic states (the "flying qubits" \cite{zol1}), which are the best long distance carriers
 of quantum information. In all experimental
 implementations of teleportation to date 
\cite{zei2,bos,braun}, and in some related proposals \cite{ral}, the stationary
qubits have been of "fleeting" optical origin.  
In earlier proposals of atomic state
teleportation  \cite{atm}, the flying qubits have been
atomic states and thereby not ideal for long distance teleportation. Our
scheme differs from these earlier 
experiments and 
proposals in using both the ideal stationary and
the ideal flying qubits. It also differs crucially from the much studied 
quantum communication 
setup in which a photon {\em directly} transfers quantum information 
from an atom trapped in a cavity to another atom in a
distant cavity \cite{zol1,zol3,Pel,van1,zolb}.
 Our scheme does not require a
 direct carrier of quantum information between
distant atoms. Joint detection of photons {\em leaking out} of 
distinct cavities enables {\em disembodied} transfer of quantum information
from an atom in one of the cavities to an atom in the other.  We thus
provide a quantum state transfer scheme that
avoids the sophisticated task of feeding
a photon into a cavity from outside \cite{zol3,Pel}.

\begin{figure}
\begin{center} 
\leavevmode 
\epsfxsize=8cm 
\epsfbox{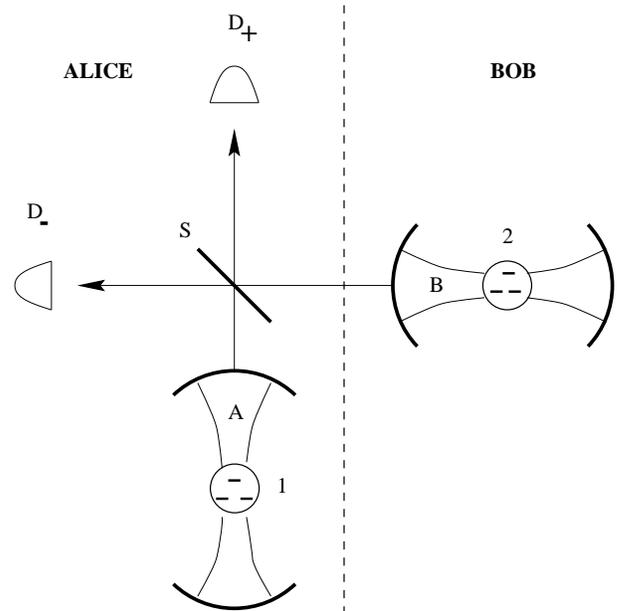}
\caption{\narrowtext The atomic state teleportation setup. 
The cavity A, atom $1$, beam splitter $S$  and the detectors $D_+$ and $D_-$ 
belong to Alice, while the cavity B and atom $2$ belong to Bob.}
\label{setup} 
\end{center}
\end{figure}

  The setup consists of two optical 
cavities, each containing a
single trapped $\Lambda$ three level atom, as shown 
in Fig.\ref{setup}. Atoms $1$ and $2$ are trapped in cavities A 
and B (supporting cavity modes A and B) respectively.The photons 
leaking out from both the cavities impinge
on the $50-50$  beam splitter $S$ and are detected at the
detectors $D_+$ and $D_-$. Initially, we assume unit efficiency
detectors (we include finite efficiency
later). The cavity A, atom $1$, beam splitter $S$ and the
detectors $D_+$ and $D_-$ belong to Alice. The cavity B with atom $2$ belongs
to Bob. We require both the cavities to be {\em one sided}
so that the only leakage of photons occur through the
sides of the cavities facing $S$. By following our teleportation protocol, Alice can teleport an
unknown state of her atom $1$ to the atom $2$ held by Bob in three
stages.

\vspace{-0.5cm}

In the preparation stage, Alice maps her atomic state to her cavity
state \cite{par}. At the same time Bob creates a maximally entangled state of his atom and his cavity mode. In the next stage (the detection stage) Alice waits
for a {\em finite} time for either or both of her detectors to click. If any
 one of the detectors register a single click during this time period, then
the protocol is successful. Otherwise Alice informs Bob about her failure.

 This protocol can be related to the standard teleportation protocol 
\cite{ben} 
by noting that the beam splitter and the
detectors constitute a device for measurement
of the joint state of the two cavities in the basis $\{|0\rangle_A
|0\rangle_B, |1\rangle_A
|1\rangle_B, \frac{1}{\sqrt{2}}(|0\rangle_A
|1\rangle_B + |1\rangle_A
|0\rangle_B), \frac{1}{\sqrt{2}}(|0\rangle_A
|1\rangle_B - |1\rangle_A
|0\rangle_B) \}$. Here $\{|0\rangle_A, |1\rangle_A \}$ and
$\{|0\rangle_B, |1\rangle_B \}$ are photon number states
in cavities A and B respectively. The teleportation is probabilistic, because it is successful
only for the pair of Bell state outcomes of the above measurement (later we
describe how to convert this to a
{\em reliable} state transfer protocol). At the end of the detection
period, if the protocol has been successful, Alice lets Bob know whether
$D_+$ or $D_-$ had clicked. This corresponds to the classical communication
part of the standard teleportation protocol \cite{ben}. Dependent on this information Bob applies a local unitary
operation to his atom to obtain the teleported state. We call this 
the post-detection stage.

   We now analyse the scheme in detail.
As we wish to look at single realizations
conditioned on detection (or not) of cavity decays, the ideal
unravelling of the system's evolution is through the quantum jump 
approach \cite{Ple2}. Let  the photon decay rate from both the cavities 
be $\kappa$. While Alice/Bob is applying a Hamiltonian $H$ to her/his atom-cavity system, its evolution 
subject to no detector click, is governed by the effective Hamiltonian (with $\hbar=1$)
$H_{\mbox{\scriptsize{eff}}}=H-i\kappa c^{\dagger}c$
(where $c^{\dagger}$ and $c$ are the creation and the destruction operators
for the cavity mode under consideration). 
The coherent evolution due to $H_{\mbox{\scriptsize{eff}}}$ is interrupted by quantum jumps when there is
a click in either the detector $D_+$ (corresponds to
an action of the operator $(c_{\mbox{\scriptsize{A}}}+c_{\mbox{\scriptsize{B}}})/\sqrt{2}$ on the joint state vector of the pair of atom-cavity systems, $c_{\mbox{\scriptsize{A}}}$ and $c_{\mbox{\scriptsize{B}}}$ being the lowering operators for modes A and
B respectively) 
or  the detector $D_-$ (corresponds to an action of the operator $(c_{\mbox{\scriptsize{A}}}-c_{\mbox{\scriptsize{B}}})/\sqrt{2}$ in the same way).

\begin{figure}
\begin{center} 
\leavevmode 
\epsfxsize=8cm 
\epsfbox{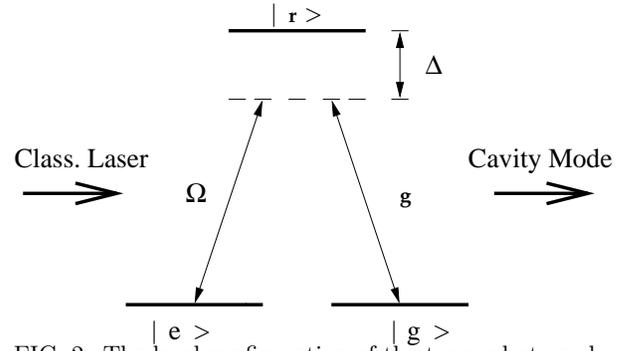}
\caption{\narrowtext The level configuration of the trapped atom showing
the fields responsible for the adiabatic evolution. The $|e\rangle \rightarrow |r\rangle$ transition being driven by
a classical laser field of coupling $\Omega$ and the $|r\rangle \rightarrow |g\rangle$ transition being driven by
the quantized cavity mode of coupling $g$. $\Delta$ is the detuning of both the 
classical laser field and the quantized field mode from their
respective transitions.}
\label{lamdatom} 
\end{center}
\end{figure}

   The three level atoms have two ground states
$|g\rangle$ and $|e\rangle$ (e.g. Zeeman sub-levels) and an excited  state $|r\rangle$ (with a spontaneous decay rate $\gamma$) as
shown in Fig.\ref{lamdatom}.
Alice and Bob use two types of time evolutions of the atom-cavity system as
their basic local operations.  The first type an adiabatic evolution (shown
in Fig.\ref{lamdatom}) which is initiated by
switching on a classical laser field which
drives the $|e\rangle \rightarrow |r\rangle$ transition with a coupling
constant $\Omega$. The $|r\rangle \rightarrow |g\rangle$ transition is driven by the quantized cavity mode of coupling $g$. Both the classical laser field
and the cavity modes are assumed to be detuned from their respective
transitions by the same amount $\Delta$. As the atom is trapped in a 
specific position 
in the cavity, we can assume that the couplings $\Omega$ and $g$ 
remain constant during
the interaction.  We choose parameters such that
 $g \Omega/\Delta^2 \ll 1 $ (the upper 
level $|r\rangle$ can then be decoupled from the evolution) and 
$\Delta \gg \gamma$ (the spontaneous
decay rate from $|r\rangle$ can be neglected). The Hamiltonian
for the evolution of the system under such conditions (and assuming $g=\Omega$ for
simplicity), is given 
by 
$H^{(1)}=E |e\rangle \langle e| + E |g \rangle \langle g|+E (c |e\rangle \langle g| +  c^{\dagger} |g \rangle \langle e|)$
where $E =g \Omega/\Delta$ \cite{Pel}.
The other local operation
accessible to Alice and Bob is the Zeeman evolution
 used to give an arbitrary phase shift of the level $|e\rangle$ relative
to the level $|g \rangle$. The Hamiltonian for this evolution
is $H^{(2)}=\delta E |e\rangle \langle e|$, where $\delta E$ is an energy
difference.

\vspace{-0.5cm}

     Let the unknown state of the atom $1$ which Alice wants to
teleport be  
\begin{equation}
|\Psi \rangle^I_1 = a|e\rangle_1 + b|g\rangle_1 ,
\end{equation}
where the superscript $I$ in $|\Psi \rangle^I_1$ stands for input and $a$ and
$b$ are complex amplitudes. We will assume that the initial state of Alice's cavity is
$|0\rangle_A$  and the initial state of Bob's atom-cavity system is
$|e\rangle_2|0\rangle_B$.
At first, Alice maps the state of atom $1$ onto the cavity
mode A by switching the Hamiltonian $H^{(1)}$ on for a period of time $t_I$ given by $\tan{\frac{\Omega_\kappa t_I}{2}}=-\frac
{\Omega_\kappa}{\kappa}$ where $\Omega_\kappa = \sqrt{4E^2-\kappa^2}$. 
Subject to no decay being recorded in the detectors,
the cavity state is given by 
\begin{equation}
|\Psi \rangle^I_A = \frac{1}{\sqrt{|a|^2\alpha^2+|b|^2}}(a \alpha |1\rangle_A + b |0\rangle_A),
\end{equation}
where $\alpha=(\frac{e^{-\frac{\kappa}{2}t_I}}{ \Omega_\kappa} 2E \sin{\frac
{\Omega_\kappa t_I}{2}})$. The probability that no photon decay takes place
during this evolution is given by $P_{ND}(A)=(|a|^2 \alpha+|b|^2)$.
Meanwhile, Bob also switches on the Hamiltonian $H^{(1)}$ in his cavity for a 
shorter length of time  $t_E$ given by $\tan{\frac{\Omega_\kappa t_E}{2}}=-\frac
{\Omega_\kappa}{2E-\kappa}$. His atom-cavity system 
 thus evolves to the entangled state
\begin{equation}
|\Psi \rangle^E_{2,B} =  \frac{1}{\sqrt{2}} (|e\rangle_2|0\rangle_B
+i|g\rangle_2|1\rangle_B).
\end{equation}
The probability that no photon decay takes place
during this evolution is given by $P_{ND}(B)=|\beta|^2$ where $\beta=\frac{e^{-\frac{\kappa}{2}t_E}}{\Omega_\kappa}2\sqrt{2} E \sin{\frac
{\Omega_\kappa t_E}{2}}$.
For simplicity, we assume that Alice and Bob 
synchronize their actions such that the preparation of the states 
$|\Psi \rangle^I_A$ and $|\Psi \rangle^E_{2,B}$ terminate at the same
instant of time. This concludes the preparation stage of the protocol.
The probability that this stage is a success is the probability
that no photon decays from either cavity during the preparation. This is given by  $P_{\mbox{\scriptsize{suc}}}(\mbox{prep})=P_{ND}(A)P_{ND}(B)$. We will
choose $\Omega_\kappa >> \kappa$ which 
makes $P_{\mbox{\scriptsize{suc}}}(\mbox{prep}) \sim 1$.

       Now comes the detection stage, in which Alice simply waits for
any one of the detectors $D_+$ or $D_-$ to click. She waits for a finite
detection time denoted by $t_D$.  Alice and Bob reject the cases in which Alice does not register any click or registers
two clicks.  The joint state of Alice's and Bob's system at the begining of the
detection stage is 
\begin{equation}
|\Phi(0)\rangle = |\Psi \rangle^I_A \otimes |\Psi \rangle^E_{2,B}.
\end{equation}
Assume Alice registers  
a single click at a time $t_j \leq t_D$. The joint state of Alice's and Bob's system
evolves as 
$|\Phi(t)\rangle_{A,2,B} = |\Psi (t)\rangle^I_A \otimes |\Psi (t) \rangle^E_{2,B}$ \cite{Ple2},
where 
$|\Psi (t)\rangle^I_A= (a\alpha e^{-\kappa t} |1\rangle_A + b |0\rangle_A)/\sqrt{|a\alpha|^2e^{-2\kappa t}+|b|^2}$
and 
$|\Psi (t) \rangle^E_{2,B}= (|e\rangle_2|0\rangle_B
+i e^{-\kappa t} |g\rangle_2|1\rangle_B)/\sqrt{1+e^{-2\kappa t}}$.
The registering of a click at one of the detectors corresponds
to the action of the jump operators $(c_{\mbox{\scriptsize{A}}}\pm c_{\mbox{\scriptsize{B}}})/\sqrt{2}$ on
the state $|\Phi(t_j)\rangle_{A,2,B}$. 
Then the resultant joint state of Alice's and Bob's system becomes
\begin{eqnarray}
|\Phi(t_j)\rangle^{J\pm}_{A,2,B}&=&  \frac{1}{\sqrt{P_{ND}(A)+2|a|^2\alpha^2 e^{-2 \kappa t_j}}}\{ (a \alpha |e\rangle_2 \nonumber
\\ &\pm& i b |g\rangle_2 )\otimes |0\rangle_A|0\rangle_B \nonumber \\
&+& e^{- \kappa t_j} a \alpha |g\rangle_2 \otimes (|1\rangle_A|0\rangle_B\pm|0\rangle_A|1\rangle_B)\}.
\end{eqnarray}
$|\Phi(t_j)\rangle^{J\pm}_{A,2,B}$ corresponds to the click being registered
in $D_{\pm}$ and the superscript $J$ stands for jump. At the end of a successful
detection stage the joint state of the cavities A, B and atom $2$ will be
$|\Phi(t_D)\rangle^{J\pm}_{A,2,B}$.
 In the post detection
stage, Bob uses $H^{(2)}$ to give $|g\rangle_2$ an extra phase shift with
respect to $|e\rangle_2$. This phase shift is $-i$ if
$D_+$ had clicked and $i$ if $D_-$ had clicked.  This concludes the entire protocol.

   We now proceed to estimate the fidelity of the teleported state generated
at Bob's end with respect to Alice's input state $|\Psi \rangle^I_1$. First we must note that though the field {\em continues to decay} even after the protocol
is over (i.e Alice has ceased to keep track of detector clicks), the reduced density matrix of atom $2$ remains unchanged,
as this atom no longer interacts with the cavity field. 
Thus the average density matrix of Bob's atom generated due to our teleportation
procedure is given by
$\rho^{Tel}_2 = \{P_{ND}(A)
|\Psi\rangle_2 \langle \Psi|_2 + 2|a|^2\alpha^2 e^{-2 \kappa t_D} |g\rangle_2 
\langle g|_2 \}/\{P_{ND}(A)+2|a|^2\alpha^2 e^{-2 \kappa t_D}\}$,
where $|\Psi\rangle_2=(a \alpha |e\rangle_2 + 
b |g\rangle_2)/\sqrt{|a|^2\alpha^2+|b|^2}$. 
The fidelity of this state with respect to the input state is 
$F(t_D,a,b)=\{P_{ND}(A)(|a|^2\alpha+|b|^2)+2|a|^2\alpha^2 e^{-2 \kappa t_D}|b|^2\}/\{P_{ND}(A)+2|a|^2\alpha^2 e^{-2 \kappa t_D}\}$.
 We see that apart from the system parameters 
$\kappa$ and $\Omega_\kappa$, the fidelity of the generated state also depends on the detection
time $t_D$ and the modulus of the amplitudes $a$ and $b$ of the initial state. It is
a teleportation protocol with a {\em state
dependent fidelity}. The fidelity does not depend on $P_{ND}(B)$ because the initial state $|\Psi \rangle^E_{2,B}$ prepared by Bob is independent
of the decay rate of his cavity.

\begin{figure}
\begin{center} 
\leavevmode 
\epsfxsize=8cm 
\epsfbox{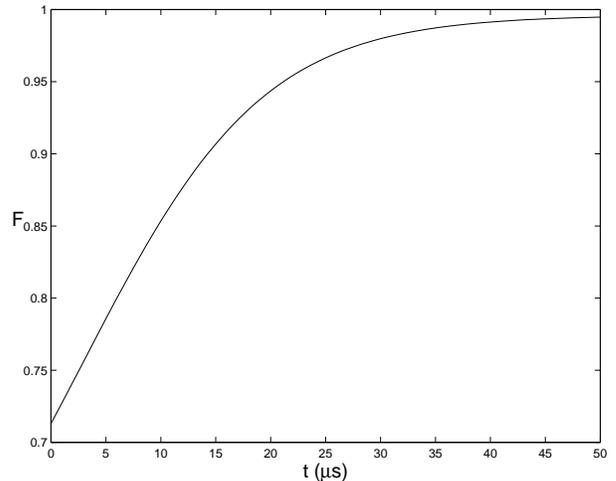}
\caption{\narrowtext The improvement of average teleportation fidelity with the length of the detection stage. The parameter regime is $(g:\Omega:\kappa:\gamma:\Delta)/2\pi=(10:10:0.01:1:100)$ 
MHz}
\label{Tfid} 
\end{center}
\end{figure}

\vspace{-0.5cm}
We plot the variation of the average fidelity of teleportation over all possible input states
as a function of the detection time $t_D$ in Fig.\ref{Tfid}. We see that
the fidelity increases with increasing detection time. This
happens because increasing the detection time decreases
the proportion of $|g\rangle_2 
\langle g|_2$ in the teleported state $\rho^{Tel}_2$ and brings it closer
to the initial state $|\Psi \rangle^I_1$ of Alice's atom. The parameter regime  used
for Fig.\ref{Tfid} \{$(g:\Omega:\kappa:\gamma:\Delta)/2\pi=(10:10:0.01:1:100)$ 
MHz\},
is carefully
chosen to satisfy all
our constraints ($g \Omega/\Delta^2 \ll 1, \Delta >> \gamma, 
\Omega_\kappa >> \kappa$).  This regime could be approached, for example, by increasing the cavity finesse of Ref.\cite{kim} by an
order of magnitude and increasing the length of that cavity to
about a millimeter while
keeping the beam waist and other parameters constant.
 As evident from
Fig.\ref{Tfid}, the average fidelity exceeds $0.99$ 
for a detection time
of about half the cavity life time.

The total probability of success
of the protocol is also state dependent and given by
$P_{\mbox{\scriptsize{suc}}}=P_{\mbox{\scriptsize{suc}}}(\mbox{prep})\times P_{1D}(0,t_D)
= (P_{ND}(A)+2|a|^2\alpha^2 e^{-2 \kappa t_D})P_{ND}(B)(1-e^{-2 \kappa t_D})/2$,
where $P_{1D}(0,t_D)$ is the probability of a single decay during the
detection period.
In the parameter domain under consideration, for $t_D=50\mu$s, we find that the
average of the probability of success over all input states is about 
$0.49$. This is a little lower than the ideal success probability of
$0.5$ (for Alice registering any of the pair of Bell state
outcomes) because the preparation stage has an extremely small, but
finite, chance of
failure.

    Let, in a real experiment, the total efficiency of photon detection (including all detector and
other unwanted losses) be $\eta$.  In the
detection stage, Alice will be able to detect only a fraction $\eta$ of all
her successful protocols. On the other hand, during this
stage, she will erroneously regard
a fraction $2 \eta (1-\eta)$ of the cases with two decays
as successful cases. Then the probability of a successful protocol changes to
$P_{\mbox{\scriptsize{suc}}}(\eta)=\eta P_{1D}(0,t_D)+
2 \eta(1-\eta) (1- P_{ND}(0,t_D)-P_{1D}(0,t_D))$, and
the fidelity of the fidelity of the teleported state would be 
$\{\eta P_{1D}(0,t_D) F(t_D,a,b) + 2 \eta(1-\eta) (1- P_{ND}(0,t_D)-P_{1D}(0,t_D))|b|^2\}/P_{\mbox{\scriptsize{suc}}}(\eta)$, where $P_{ND}(0,t_D)$ is 
the probability of no decay during the
detection period. In the parameter domain under consideration, and for $\eta$
not lower than $0.1$,
we can neglect the effect of undetected photon losses during the preparation stage on the fidelity. 
For a $\eta$ of $0.6$ and detection times large compared
to the cavity decay time, the fidelity of the state 
at Bob's end becomes $\sim 0.81$.
  
     The main practical role of teleportation 
is to act as a device to link up distant quantum processors with
entanglement. To set up entanglement between their atoms, Alice and Bob must
both prepare their respective atom-cavity systems in the state
$(|e\rangle|0\rangle
+i|g\rangle|1\rangle)/\sqrt{2}$ during the preparation stage.
Entanglement between the atoms is established if there is a single click during
the detection period. The resultant entangled state is
$|\Psi_{12}\rangle = \{\eta(1-e^{-4\kappa t})/4\} |\psi^{\pm}\rangle \langle \psi^{\pm}| + \{\eta(1-\eta)(1+e^{-4\kappa t}-2 e^{-2\kappa t})/2\} |g\rangle_1  |g\rangle_2 \langle g|_1 \langle g|_2$, where $|\psi^{\pm}\rangle = |e\rangle_1  |g\rangle_2 \pm |g\rangle_1  |e\rangle_2$. 
The relative entropy of entanglement of this state can be calculated \cite{vlat}
and for $t_D$ large compared to the cavity decay time and a reasonable $\eta$ of $0.6$ it is about $0.16$, while for
a high $\eta$ of $0.9$, it is about $0.48$ (note that $|\Psi_{12}\rangle$
is entangled for arbitrary $\eta$). From the viewpoint of setting up of
entanglement, our scheme is rather close to the scheme
described by Cabrillo {\em et al.} \cite{cab}. But the efficiency of
success can be much higher (nearing $0.5$).

     The above probabilistic teleportation
protocol can be modified to {\em teleportation
with insurance}, so that in the cases when the protocol is
unsuccessful, the original state of Alice's atom $1$ is not destroyed,
but mapped onto another reserve atom $r$ trapped in Alice's cavity.  
To accomplish this, Alice has to follow the {\em local redundant
encoding} of Ref.\cite{van1} and codes her initial state $|\Psi \rangle^I_1$
as $a (|e\rangle_1 |g\rangle_r + |g\rangle_1 |e\rangle_r) +
b (|g\rangle_1 |g\rangle_r + |e\rangle_1 |e\rangle_r)$. After this, she
just follows the same protocol as before. But in cases when the protocol
is unsuccessful, she is left with either 
the state $a |g\rangle_r + b |e\rangle_r$
or a state that can be converted to $a |g\rangle_r + b |e\rangle_r$ by a
known unitary transformation. She can now exchange the roles of atom $1$
and atom $r$ and try to teleport the state $|\Psi \rangle^I_1$ again.
She can repeat this procedure until teleportation
is successful (Of course, this holds true perfectly only when $\eta=1$).
 
 To conclude, we have presented a simple
scheme for atomic state teleportation, which could be implemented 
by trapping single three level atoms in a cavity. Moreover,
by adding one more atom to Alice's cavity, it can be converted to
a {\em reliable} state transfer protocol. This state transfer protocol
can be viewed as an {\em alternative} to designer laser pulses for
 transferring (Refs.\cite{zol3,Pel}) 
quantum information
into a cavity from outside. This state transfer should work for
distances of the order of the absorption length scales of a fibre. 
The model independent portions of the 
analysis of communication
through a noisy 
quantum channel \cite{van1,zolb,brig} should
carry over to this decay-induced
scenario of state transfer.  
The scheme described here is also a 
rare example of a quantum decay playing a constructive role in quantum
information processing.
   
 We thank  S. L. Braunstein, W. Lange, E. Paspalakis and S. J. van Enk for 
valuable discussions. Part of this work was carried out
during the Complexity, Computation and Physics of Information
Programme of the Isaac Newton Institute, and the European Science Foundation
QIT-Programme on Quantum Information Theory. This work was supported in part by the 
Inlaks Foundation, the UK Engineering and Physical Sciences Research Council, the European Union, the Leverhume Trust, Elsag-Bailey and 
Hewlett-Packard.

\end{multicols}
\end{document}